# The equation of state for non-ideal quark gluon plasma

N.M. El Naggar[(1)], L.I. Abou Salem[(1)], A.G. Shalaby[(1)] and M. A. Bourham[(2)],

(1) Physics Department, Faculty of Science, Benha University, Egypt
(2) Nuclear Engineering Department, North Carolina State University, USA

## Abstract

The mass spectra of quarkonium systems at T= 0 are analyzed by solving the non-relativistic radial wave equation using the internal energy potential. The QGP matter is studied through the dissociations of quarkonium systems. A modified form of the internal energy potential function is used to determine the EoS at different number of quark flavors by using Mayer's cluster expansion theory and phenomenological thermodynamic model. The thermodynamic model gives a good agreement with the lattice results rather than Mayer's cluster expansion theory. One can conclude that, the Mayer's cluster expansion theory may be more suitable to study a weakly coupled plasma while, the QGP may be considered as a strongly interacting plasma.

## I. Introduction

In the heavy ion collison, there are many signature of deconfinment state creation; one of these signatures is the anomalous suppression of heavy quarkonium production. The heavy mesons produced before the creation formation of a thermalized quark-gluon-plasma and tend to dissociate in the deconfined state. This phenomenon can be described by the screening of the quark-antiquark ($q\bar{q}$) interaction by the large number of color charges in the medium. This mechanism is similar to the Debye screening by electromagnetic charges in the quantum electrodynamics (QED) plasmas [1]. The suppression of heavy quarkonium at finite temperature greater than zero, T ≥ 0 concerning to the quantum chromodynamics (QCD) has been studied [2]. So that, the dissociation temperature of a particular quarknoium, playes an important role to understand the mechanism of quarkonium dissociation (deconfinemt) in the quark-gluon-plasma (QGP). Brambilla et al.,[3] have shown that the heavy quark antiquark potential at finite temperature develops an imaginary part that is responsible of the quarkonium dissociation in medium.

Emails : [1] loutfy.abousalem@fsc.bu.edu.eg

[1] asmaa.shalaby@fsc.bu.edu.eg, asmaa.gaber76@gmail.com; [2] bourham@ncsu.edu

The mass spectra at T= 0, can be reproduced by using the potential models. So that, the lattice QCD simulations could find a relevant potential term at T > 0. This can be done by studying the free energy between a static $q\bar{q}$ pair at finite temperature [4, 5]. From the lattice QCD calculations critical temperature; $T_c$ can be determined in which the confining part of the free energy has no effect and vanishes [6].

Then the free energy obtained in these calculations can be used to establish the convenient potential model at T > 0. However, in other works [7-10], the internal energy can be used as a potential energy. The explansion of how the different potential models be applied to quarkonium states at temperature greater than zero (i.e T > 0) is still not completely clarified. Till now, it is believed quantum chromodynamics (QCD) at high temperature to be in a quark gluon plasma (QGP) phase, where color charges may be screened rather than confined [11]. This implies that, at high energy density $\varepsilon$ or baryon density $\rho$, hadron state goes to deconfined state known as the QGP [12].

In the recent years, there was a lot of theoretical experimental, and lattice calculation of QCD results. Although there is no confirmed evidence for QGP creation and the order of the phase transition. There is a large amount of study attempts to explain such a matter and the EoS using different models [13-19]. Recently, the Generalized Uncertainty Principle (GUP), used to derive the thermodynamics of ideal Quark-Gluon Plasma (QGP) at a vanishing chemical potential [20].

The EoS can be applied directly to study the dynamical quark-gluon plasma (QGP), even in case of interpretation of the heavy-ion experiments or in the framework of the theoretical modeling to study the behavior of hot and dense matter in the early universe [21]. Liu, Shen and Chiang [22] are used the Cornell potential in the approach of Mayer's cluster expansion to calculate the EoS and the energy density of the QGP.

In the work we discuss a quantity of interest, the plasma parameter, $\Gamma$, which can be defined as the ratio of the potential energy to the kinetic energy. In QED plasma (classical plasma), the parameter has four different regimes: weakly coupled or gas regime for $\Gamma \prec 1$, liquid regime for $\Gamma \approx$ 1-10, glassy liquid regime for $\Gamma \approx$ 10-100, and solid regime for $\Gamma$ > 300 [23]. The plasma parameter $\Gamma$ is defined as the ratio of potential energy to the kinetic energy $\Gamma\ =<PE>/<KE>$.

Strongly coupled plasma (SCP) is defined as a plasma in which the plasma parameter is of unit 1 or greater and the Boltzmann distribution for electrons and ions is given by $n_e = n_0\, e^{\frac{e\phi}{kT_e}}$ and $n_i = n_0\, e^{\frac{-e\phi}{kT_i}}$, respectively [23-28]. This parameter $\Gamma$ is used as a measure of the interaction strength in EM plasmas.

## II. The Bound State Problem

The bound state energies of heavy quarkonium systems ($c\bar{c}$) and ($b\bar{b}$) are calculated at different temperatures using the non-relativistic radial wave equation which given as;

$$[\frac{d^2}{dr^2} - \frac{l(l+1)}{r^2} - \frac{2\mu}{\hbar^2} V(r,T) + K^2]\, \phi_l(r) = 0 \qquad (1)$$

Where, $l$ is the orbital quantum number, $K^2 = \frac{2\mu E}{\hbar^2}$ and $\phi_l(r)$ is the radial wave function. The boundary conditions are given as;

$$\phi_l(0) = \phi_l(\infty) = 0 \qquad (2)$$

$V(r, T)$ can be taken as the internal energy potential $U_1(r, T)$,

$$U_1(r,T) = F_1(r,T) - T \cdot \frac{\partial F_1(r,T)}{\partial T} \qquad (3)$$

Where the free energy potential, $F_1(r, T) = -\frac{4}{3}\frac{\alpha_s}{r} e^{-m_D(T).r}$. And the function $e^{-m_D(T).r}$ is the screening term [10].

In the present work, the free energy function $F_1(r, T)$ can be modified through adding a linear term; due to the confinement behavior of the strong interactions at large distances;

$$F_1(r,T) = -\left(\frac{4}{3}\frac{\alpha_s(T)}{r} - C\,r\right) e^{-m_D(T).r} \qquad (4)$$

Where, C is a free parameter, and $\alpha_s(T)$ is the running coupling constant which is given by;

$$\alpha_s(T) = \frac{2\pi}{(11-\frac{2}{3}n_f)\,\ell n(\frac{T}{\Lambda_\sigma})} \tag{5}$$

Where, $n_f$ is the number of quark flavors, ($n_f$ = 0, 2, 3). From lattice QCD computations [10], the parameter $\Lambda_\sigma = \beta\,T_c$ where, $\beta = 0.104 \pm 0.009$.

In the present work, $T_c$ is taken as, $T_c = 0.2$ GeV. The Debye screening mass $m_D(T)$ [29] is given by [10],

$$m_D(T) = \gamma . \alpha_s(T) . T, \tag{6}$$

Where, $\gamma = 4\pi\eta c_\sigma$. In table (1) all the parameters used are listed.

**Table (1) The parameters of the internal energy**

| Parameter | Value | Ref. |
|---|---|---|
| $\beta$ | $0.104 \pm 0.009$ | [4,10] |
| $c_\sigma$ | $0.566 \pm 0.013$ | [4, 10] |
| $\eta$ | 2.06 | [4, 10] |
| $T_c$ | 0.2 GeV | Present work |
| C | $0.135 \pm 0.015$ GeV$^2$ | Present work |
| $m_c$ | $1.361 \pm 0.022$ GeV | Present work |
| $m_b$ | $4.694 \pm 0.063$ GeV | Present work |

According to eqs. (3, 4), the internal energy potential can be rewritten as,

$$U_1(r,T) = -\left(\frac{4}{3}\frac{\alpha_s(T)}{r} - Cr\right)e^{-m_D(T).r} - \frac{4}{3}T\left[\left(\gamma.\frac{A_1}{A_2}\left(\alpha_s(T) - \frac{A_1}{A_2^2}\right) + \frac{A_1}{TA_2^2}.\frac{1}{r}\right)\right]e^{-m_D(T).r}$$

$$+ C.T.r^2.\gamma\left(\alpha_s(T) - \frac{A_1}{A_2^2}\right)e^{-m_D(T).r}$$

(7)

For simplicity, $A_1 = \dfrac{2\pi}{(11-\frac{2}{3}n_f)}$ and $A_2 = \ell n(\dfrac{T}{\Lambda_\sigma})$

The total mass of the different quarkonium states (resonance masses) is given by;

$$M_{nl} = 2m_q + \varepsilon_{nl} \qquad (8)$$

Where, $m_q$ is the quark mass, and $\varepsilon_{nl}$ is the "binding energy". Equation (1) can be re-written as,

$$\frac{d^2\phi_l(r)}{dr^2} + [\lambda - C(r,T)]\phi_l(r) = 0 \qquad (9)$$

Where, $\hbar = 1$, $\lambda \equiv K^2 = \frac{2\mu E}{\hbar^2}$ and $C(r, T) = 2\mu V(r,T) + \frac{l(l+1)}{r^2}$

Introducing the dimensionless variables t and $\rho_l(t)$ [30, 31] in equation (9) where,

$$t = \frac{1}{1+\frac{r}{r_0}} \quad \text{and} \quad \rho_l(t) = t\,\phi_l(t) \quad \text{Where, } r_0 = 1 \text{ Gev}^{-1} \text{ one gets,}$$

$$\frac{d^2\rho_l}{dt^2} + \frac{r_0^2}{t^4}[\lambda - C(t,T)]\rho_l(t) = 0 \qquad (10)$$

With the boundary conditions,

$$\rho_l(0) = 1 \quad , \quad \rho_l(1) = 0 \qquad (11)$$

To transform eq. (10) to a true eigen-value equation, the range of t from (0, 1) can be divided into (n+2) points with the interval h and labeled with subscript j and the boundary conditions (11) at $j = 0$ and n+1 can be written as,

$$\rho_0 = \rho_{n+1} = 1 \qquad (12)$$

Using the finite difference approximation [32],

$$\frac{d^2\rho}{dt^2} = \frac{1}{12h^2}(-\rho_{j-2} + 16\rho_{j-1} - 30\rho_j + 16\rho_{j+1} - \rho_{j+2}) + O(h^4) \qquad (13)$$

Substitute into eq. (10) one gets,

$$(-\rho_{j-2} + 16\rho_{j-1} - 30\rho_j + 16\rho_{j+1} - \rho_{j+2}) + \frac{12h^2}{(jh)^4}[C(jh,T) - \lambda]\rho_j = 0 \qquad (14)$$

Equation (14) is a set of linear equations in $\rho_j$ and can be written in the matrix form,

$$S\rho = 0 \qquad (15)$$

Where S is a (n× n) symmetric matrix and $\rho$ is n-dimensional column matrix. Eq. (15) can be transformed to a true eigen-value equation and solved numerically by using Jacobi method [32, 33].

Table (2) is a list of the resonance masses $M_{nl}$ in (GeV) of $c\bar{c}$ and $b\bar{b}$ states. We have calculated them by solving the Schrödinger equation numerically by using the internal energy potential at T = 0. In table (2) the calculated masses of $c\bar{c}$ and $b\bar{b}$ states according to different previous potential forms and the internal energy potential are given. One can see that the masses calculated by using the internal energy potential are very close to the experimental data.

**Table (2):** The mass spectra of $c\bar{c}$ and $b\bar{b}$ bound states by using the internal energy potential compared to the experimental masses and other theoretical potentials.

| | $nl$ | State (GeV) $M_{nl}$ [22, 34,35] | The present work | Cornell potential [22] | Phenomenological potential [34] |
|---|---|---|---|---|---|
| $c\bar{c}$ | 1S | $J/\psi$ (3.097 ± 0.001) | 3.097 | 3.0697 | 3.097 |
| | 2S | $\psi'$ (3.686 ± 0.0027) | 3.687 | 3.6978 | 3.684 |
| | 3S | $\psi''$ (4.040 ± 0.0027) | 4.047 | 4.1696 | 4.096 |
| | 4S | $\psi$ (4.415 ± 0.0062) | 4.415 | - | 4.427 |
| | 1P | $\chi_c$ (3.506 ± 0.0041) | 3.500 | 3.5003 | 3.520 |
| | 1D | $\psi$ (3.768 ± 0.0036) | 3.769 | - | 3.671 |
| | 2D | $\psi$ (4.159 ± 0.02) | 4.134 | - | 4.076 |
| | $nl$ | State (GeV) $M_{nl}$ [22, 34,35] | The present work | Cornell potential [22] | Screened potential [35] |
| $b\bar{b}$ | 1S | $\Upsilon$ (9.460 ± 0.00026) | 9.460 | 9.4450 | 9.460 |
| | 2S | $\Upsilon'$ (10.0233 ± 0.00031) | 10.0227 | 10.0040 | 10.016 |
| | 3S | $\Upsilon''$ (10.3553 ± 0.0005) | 10.3551 | 10.3547 | 10.351 |
| | 4S | $\Upsilon$ (10.580 ± 1.0002) | 10.580 | - | 10.611 |
| | 5S | $\Upsilon$(10.865 ± 0.0008) | 10.7808 | - | 10.831 |
| | 1P | $\chi_b$ (9.9002 ± 0.00026) | 9.9003 | 9.8974 | 9.918 |
| | 2P | $\chi_b$ (10.268 ± 0.00022) | 10.2522 | - | 10.269 |
| | 1D | $\psi$ (10.161 ± 0.0006) | 10.1557 | - | 10.156 |

### III. The QGP equation of state by using Mayer's cluster expansion theory

Mayer's theory of plasma is described in [36, 37]. The EoS is one of the most basic information in the case of studying the QGP matter;

$$\frac{P}{T} = n_q + n_{\bar{q}} + S - n_q \frac{\partial S}{\partial n_q} - n_{\bar{q}} \frac{\partial S}{\partial n_{\bar{q}}} \tag{16}$$

Where P, T, $n_q$, $n_{\bar{q}}$ are the pressure, the temperature, the densities of the quarks and the antiquarks, respectively. The entropy S is given by [36],

$$S = \int dI \sum_{\nu \geq 1} \frac{1}{16\pi^3} (-k^2)^\nu V_\ell^{\nu+1} , \tag{17}$$

Where, $k^2 = \frac{n_q + n_{\bar{q}}}{T}$, $V_\ell$ is the interaction potential in the momentum space and

$dI = 4\pi \ell^2 d\ell$, therefore,

$$\frac{\partial S}{\partial k^2} = \frac{k^2}{4\pi^2} \int_0^\infty d\ell \frac{V_\ell^2}{1+k^2 V_\ell} \tag{18}$$

Therefore equation (16) can be written as;

$$\frac{P}{T} = k^2 T + S - k^2 \frac{\partial S}{\partial k^2} \tag{19}$$

So, the internal energy potential eq. (7) can be transformed by Fourier transformation to the momentum space and rewritten as,

$$U_1(\ell, T) \frac{-16\pi}{3} \frac{\alpha_s(T)}{\ell^2 + m_D^2(T)} + \frac{16\pi C m_D(T)}{(\ell^2 + m_D^2(T))^2} =$$

$$-\frac{32}{3}\pi T \gamma \alpha_s(T).m_D(T) \left( \frac{-A_1 + A_1.A_2}{A_2^2} \right) \frac{1}{m_D^2 + \ell^2}$$

$$-\frac{16\pi}{\ell^2 + m_D^2(T)} \left( \frac{A_1}{A_2^2} \right) + 96\pi C T \gamma m_D(T). \left( \frac{-A_1 + A_1.A_2}{A_2^2} \right) . \left( \frac{m_D^2(T) - \ell^2}{\left(\ell^2 + m_D^2(T)\right)^2} \right)$$

$$\tag{20}$$

The energy density can be calculated by the following relation [10]

$$\varepsilon = T\frac{\partial P}{\partial T} - P \tag{21}$$

Taking $k^2 \approx a_f T^2$ in which, $a_f$ is the Stefan-Boltezmann constant [38] is given by,

$$a_f = (16 + \frac{21}{2} n_f)\frac{\pi^2}{90} \tag{22}$$

## IV. The phenomenological thermodynamic model

The EoS of SCP as a function of $\Gamma$ is given as [24],

$$\varepsilon_{QED} = \left(\frac{3}{2} + U_{ex}(\Gamma)\right)nT \tag{23}$$

Where, $U_{ex}(\Gamma)$ is the non-ideal or excess contribution to EoS and is given as [24];

$$U_{ex}(\Gamma) = \frac{U_{ex}^{Abe}(\Gamma) + 3\times 10^3 \Gamma^{5.7} U_{ex}^{OCP}(\Gamma)}{1 + 3\times 10^3 \Gamma^{5.7}} \tag{24}$$

Where the functions of $U_{ex}^{Abe}(\Gamma)$, $U_{ex}^{OCP}(\Gamma)$ are given as [24];

$$U_{ex}^{Abe}(\Gamma) = \frac{-\sqrt{2}}{2}\Gamma^{3/2} - 3\Gamma^3[\frac{3}{8}\ln(3\Gamma) + \frac{\gamma}{2} - \frac{1}{3}] \tag{25}$$

$$U_{ex}^{OCP}(\Gamma) = -0.898004\ \Gamma + 0.96786\ \Gamma^{1/4} + 0.220703\ \Gamma^{-1/4} - 0.86097 \tag{26}$$

The term $U_{ex}^{Abe}$ was derived by Abe [39] and is valid for $\Gamma < 0.1$, and $\gamma = 0.57721...$ is the Euler's constant. The term $U_{ex}^{OCP}$ determined by simulation of one component plasma (OCP). The OCP is occurred when a single species of charged particles distributed in a uniform background of neutral charges, and is valid for $1 \leq \Gamma < 180$. Considering the model proposed by Bannur [24] that the hadron (confined state) exists at $T < T_c$ and goes to QGP (deconfined state) at $T > T_c$. In ref. [24] the plasma parameter $\Gamma$ is determined for the Coulomb potential.

$$\Gamma \equiv \frac{\langle PE \rangle}{\langle KE \rangle} = \frac{\frac{4\alpha_s}{3r_{av}}}{T} \tag{27}$$

The coupling constant $\alpha_s \approx 0.5$, $r_{av} \approx 1\,\text{fm}$, $r_{av} = \left(\frac{3}{4\pi n}\right)^{1/3}$, where "n" is the density.

In the present work the plasma parameter $\Gamma$ is calculated quantum mechanically as,

$$\langle Q \rangle = \int \psi^* \hat{Q} \psi \, d\tau$$

In which we have used the wave function (eigen-function) that produced for the calculation of the bound state energies (eigen-values). For the SCQGP model of eq. (23) to include the relativistic quantum effects as indicated in ref. [24]. Hence, eq. (23) can be re-written as,

$$\varepsilon = (2.7 + U_{ex}(\Gamma)) \, n \, T \tag{28}$$

Where the first term (2.7 n T) corresponds to the ideal EoS, which may be written as, $\varepsilon_s = 3 a_f T^4$.

$$\frac{\varepsilon_s}{n} = \frac{3 a_f T^4}{1.1 a_f T^3} = 2.7 \, T \tag{29}$$

One can calculate the expectation value of the internal energy potential from the wave function reproduced from solving of the Schrödinger. From eq. (29) one obtains e($\Gamma$) the energy density normalized to the ideal one:

$$e(\Gamma) \equiv \frac{\varepsilon}{\varepsilon_s} = 1 + \frac{1}{2.7} U_{ex}(\Gamma) \tag{30}$$

From eq.(21) and eq. (28) one can get the EoS (pressure) as following,

$$\frac{P}{T^4} = \left( \frac{P_0}{T_0} + 3 a_f \int_{T_0}^{T} d\tau \, \tau^2 e(\Gamma) \right) \tag{31}$$

Where $P_0$, is the pressure at temperature $T_0$ and may be taken from one of the lattice data points or at critical temperature $T_c$.

## V. Results and discussion

In figure (1) the internal energy potential $U_1(r, T)$ eq. (7) versus r is plotted at T = 0 and compared with the Cornell potential [22]. Also, it is plotted at different temperature values, T= (0.5-1.5)$T_c$. One can notice that, at small r both potentials behave similarly approximately and intersect at $r \approx 1$ GeV$^{-1}$ in which the Coulomb term is more effective. At large distances the behavior of both potential forms is completely different where the confinement term of the potential is more effective. Also, the behavior of the deconfinement mechanism at large separation (r) and high temperature (T > $T_c$) is shown.

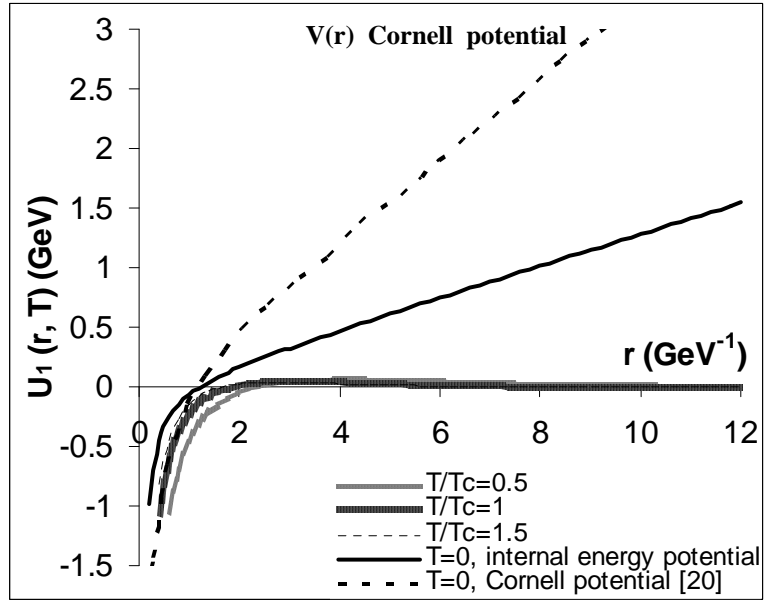

**Fig. (1) The internal energy potential $U_1(r, T)$ versus r at different temperature T= (0.5-1.5)$T_c$ and at T = 0.**

In figure (2) the running coupling constant $\alpha_s(T)$ versus temperature T is plotted at different number of quark flavors. One can see that, the running coupling decreases logarithmically as T increases for different number of flavors ($n_f$ = 0, 2, 3). The behavior of the Debye screening mass at different temperature is studied [see figure (3)]. This figure shows the calculated values of the Debye screening mass; $m_D$; versus T/$T_c$ and the lattice results [10, 29]. In this case the results predict that; $m_D(T) \propto \alpha_s(T).T$, instead of the usual dependence $\sqrt{\alpha_s(T)}$ T [10].

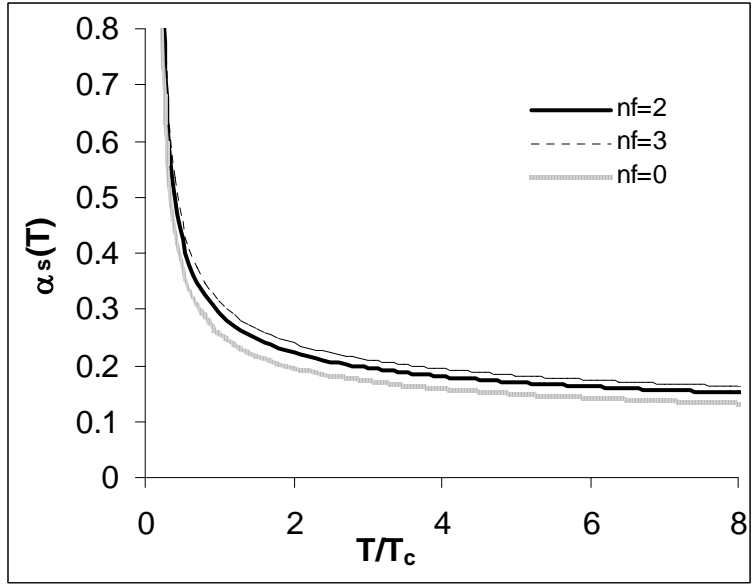

**Fig. (2) The running coupling constant $\alpha_S(T)$ versus T/T$_c$ at different number o quark f flavors n$_f$ =0, 2, 3.**

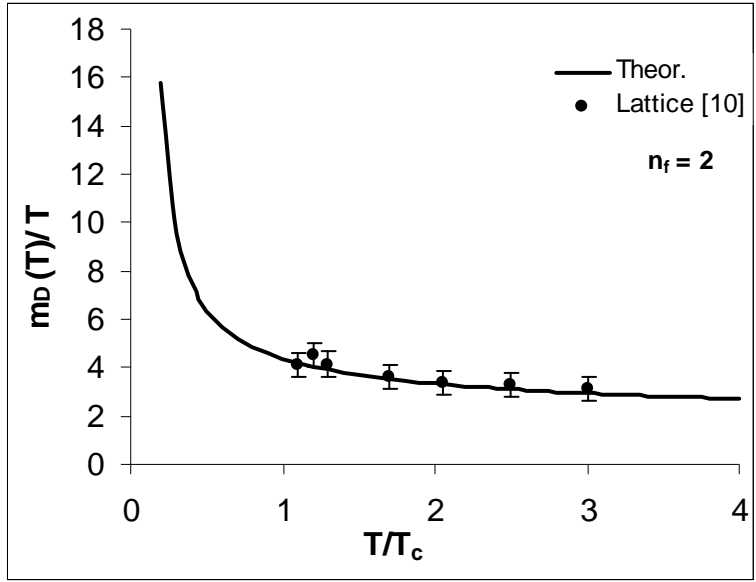

**Fig. (3) The Debye screening mass m$_D$(T)/T versus T/T$_c$ compared with the lattice data [10].**

Figure (4) shows the theoretical calculation of the EoS using Mayer's expansion theory; from eq. (19) at different number of quark flavors n$_f$ = 0, 2, 3.

In the present calculations the critical temperature is taken as, $T_c = 0.2$ GeV. The solid line and the different dashed lines represent the theoretical calculations by using Mayer's cluster expansion theory and the symbols are the lattice results. It is clear that, a suitable qualitative agreement between the theoretical calculation and the lattice results especially at the intermediate temperature range at $n_f = 0$. While at $n_f = 2$, it is clear that the present theoretical calculation does not match the lattice results. However, at $n_f = 3$ a qualitative agreement between the present calculations and the lattice results are obtained.

Generally one can conclude that, the Mayer's cluster expansion theory is more suitable to study a weakly coupled plasma, while the QGP may be strongly coupled plasma [14].

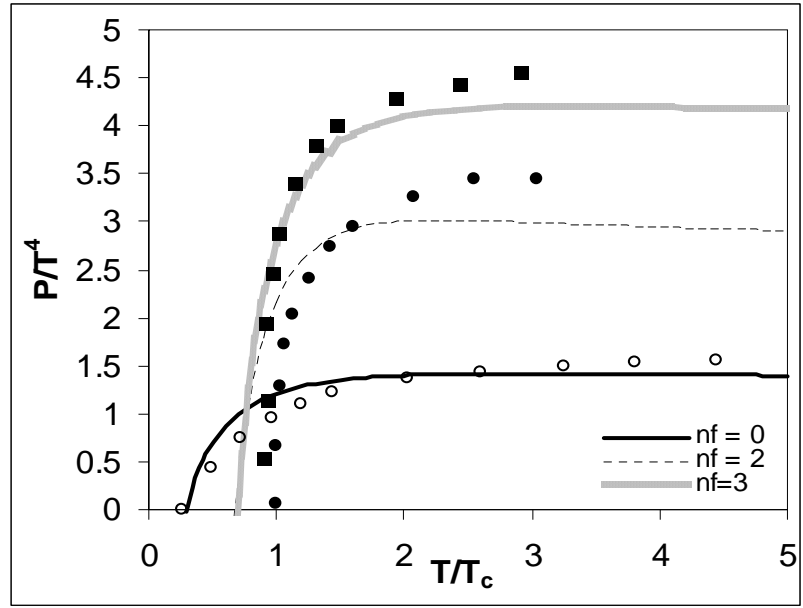

**Fig. (4) The equation of state $P/T^4$ versus $T/T_c$, $T_c$= 200 MeV, (solid and dashed lines) are the theoretical calculations and (symbols) are the lattice results [24, 40].**

Figure (5) shows the calculated energy density $\mathcal{E}/T^4$ by using Mayer's cluster expansion theory at different number of quark flavors, $n_f = 0, 2, 3$ versus $T/T_c$. The solid and dashed lines are the theoretical calculations and the symbols are the lattice results.

From this fig. one can see that, at $n_f = 0$, a qualitative agreement between the present calculations and the lattice results is obtained at high temperatures, but at $T < 2T_c$ the theoretical calculation does not match the lattice results. While at $n_f = 2$ it is clear that, the theoretical calculations doesnot give agreement with the lattice results at small temperature range. But at $n_f = 3$ it is clear that, the theoretical calculation matches with the lattice results especially at high temperature range.

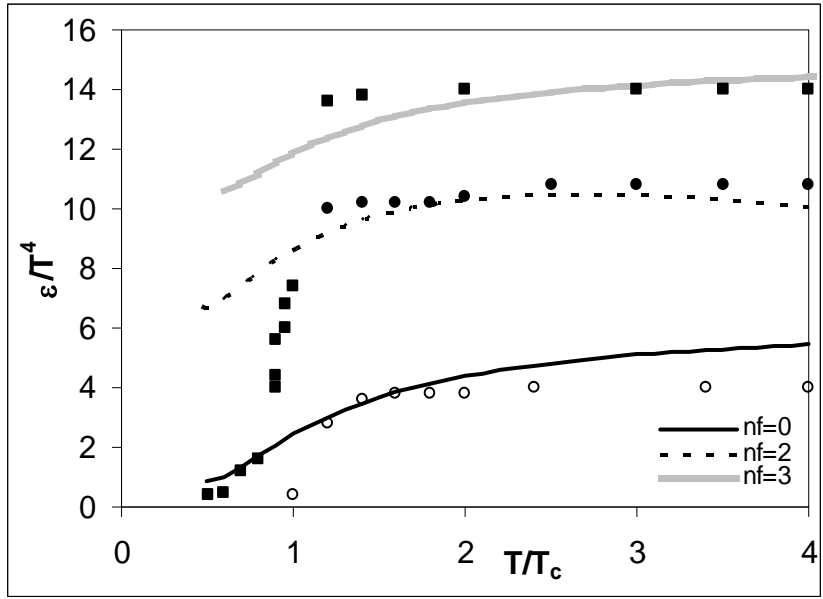

**Fig. (5) The equation of state $\varepsilon/T^4$ versus $T/T_c$, $T_c = 200$ MeV, (solid & dashed lines) are the theoretical calculations and (symbols) are the lattice results [22, 40].**

In the present work the plasma parameter is calculated quantum mechanically using the wave function produced from the numerical solution of the Schrödinger equation for quarkonium bound state system. Figure (6) shows the behavior of the calculated plasma parameter $\Gamma(T)$ using the internal energy potential. It is clear that, the largest value at $T/T_c \approx 1$, and tends to zero at very high temperature $T/T_c \approx 5$.

Equations (31) are used to calculate the EoS by using a phenomenological thermodynamic model. Figure (7) shows the present calculations of the equation of state (EoS) versus $T/T_c$ at different number of quark flavors ($n_f = 0, 2, 3$) in comparison with the theoretical calculations of the Cornell potential [24] and the lattice results.

From this figure, one can see that, the present calculations of the EoS using the internal energy potential give more agreement with the lattice results than the Cornell potential calculations at $n_f = 0, 3$. But at $n_f = 2$ it is clear that, the Cornell potential calculations give more agreement in this case with the lattice results.

The behavior of the energy density $\varepsilon/T^4$ versus $T/T_c$ is calculated by using equation (30), at different number of quark flavors, $n_f = (0, 2, 3)$.

In figure (8), the solid lines represent the present calculations of the EoS by using the internal energy potential, the dashed lines are the theoretical calculations by using Cornell potential [24], and the symbols are the lattice results [24, 41, 42].

From figure (8) one can see that, the present calculations using the modified internal energy potential give a satisfied agreement at all temperature range with the lattice results compared with the Cornell potential calculations, especially at $n_f = 0, 3$. While at $n_f = 2$ the Cornell potential calculations give a slight better fit with the lattice results.

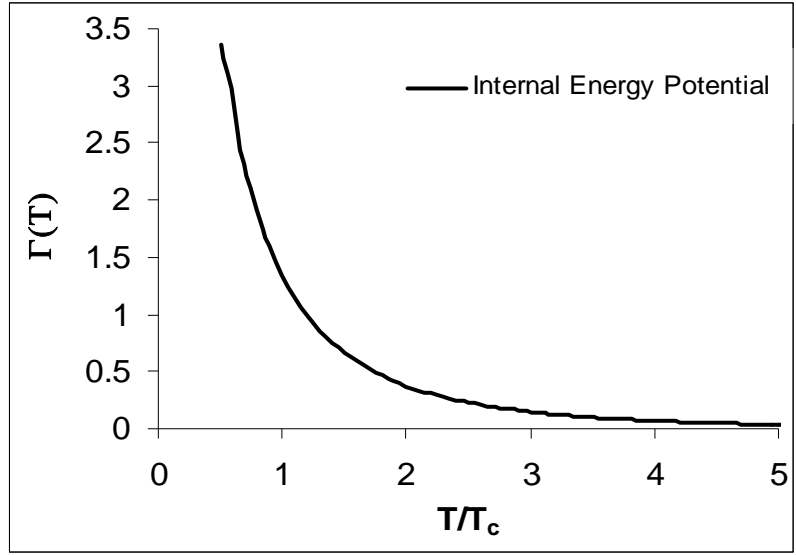

**Fig. (6) The plasma parameter $\Gamma(T)$ versus $T/T_c$ calculated by the internal energy potential.**

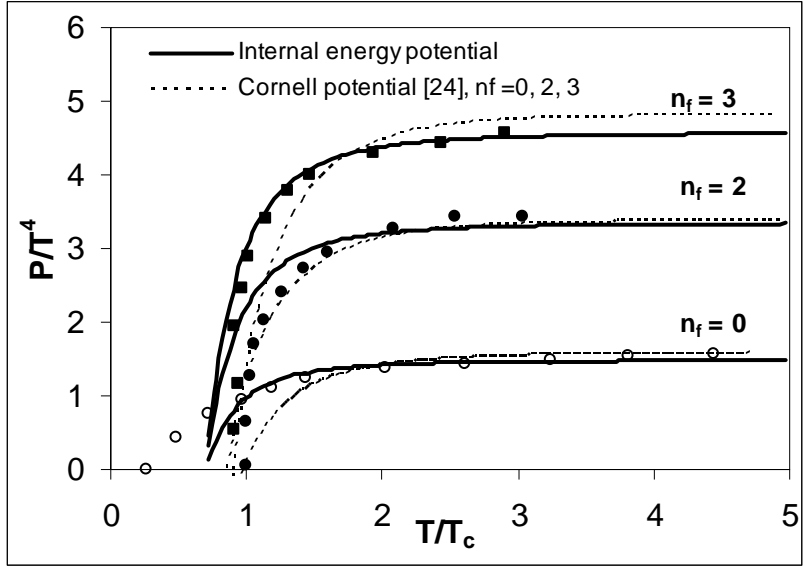

**Fig. (7)** The equation of state $P/T^4$ versus $T/T_c$, $T_c = 200$ MeV, (solid lines) are the theoretical calculations, dashed lines are the EoS calculated by Cornell potential and (symbols) are the lattice results [24,40].

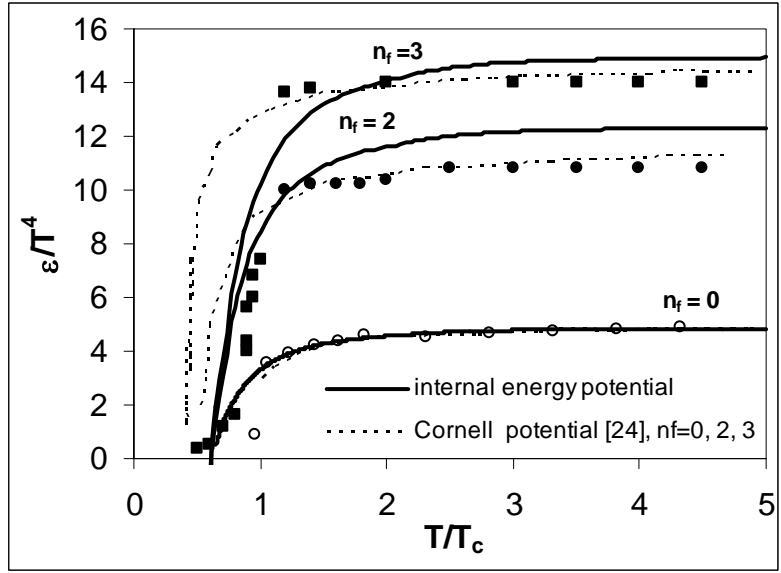

**Fig. (8)** The energy density $\varepsilon/T^4$ versus $T/T_c$, $T_c = 200$ MeV, (solid lines) are the theoretical calculations, dashed lines are the EoS calculated by Cornell potential and (symbols) are the lattice results [24,40].

Once the pressure (P) and the energy density ($\varepsilon$) are calculated, one can calculate the trace anomaly or the interaction measure quantity ($\Delta$), which is one of the most important quantities in the studying of the quark-gluon plasma phase transition.

In figure (9) the interaction measure; $\Delta = \dfrac{\varepsilon - 3P}{T^4}$; is plotted versus $T/T_c$ with the lattice results [24, 41]. In this figure we have calculated the deviation between the energy density $\varepsilon$ of the QGP system and the corresponding one in case of the ideal gas plasma ($P = \dfrac{1}{3}\varepsilon$). From this figure one can see that, $\Delta$ tends to zero for large T and $\Delta$ still has value and does not vanish up to $T \approx 3T_c$ [2, 42].

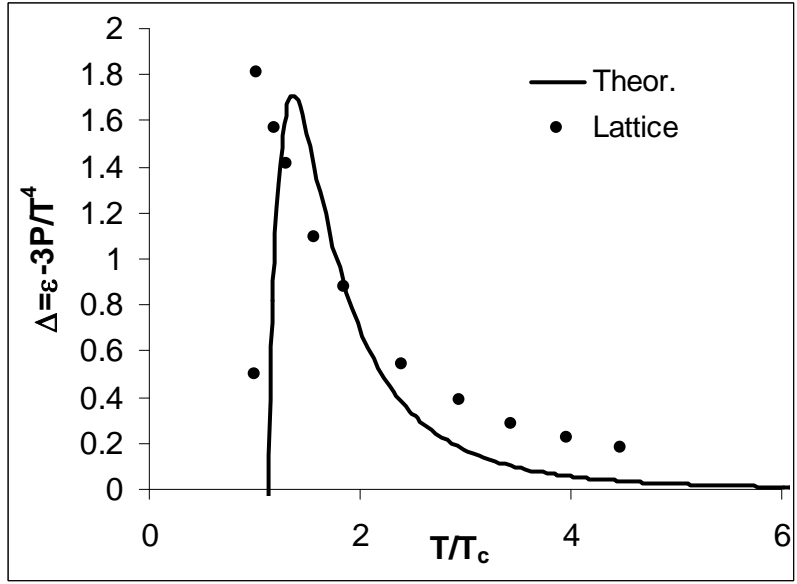

**Fig. (9) The deviation $\Delta = \dfrac{\varepsilon - 3P}{T^4}$ versus $T/T_c$, the solid line is the theoretical calculations and the symbols are the lattice results [41].**

### Conclusion

In this work, we introduce a linear term to the internal energy potential function. This modification provided a linear part within the effect of the screening term in both parts of the free energy function. Then we have studied the applicability of using this potential form to the dissociations of $c\bar{c}$ and $b\bar{b}$ systems, and the study of the equation of state for such matter. From these calculations, the Mayer's cluster expansion theory has shown poor fit with the lattice results. So that, Mayer's cluster expansion may be more suitable to study a weakly coupled plasma while the QGP may be considered as a strongly interacting plasma.

The thermodynamic model calculations depending on the plasma parameter $\Gamma(T)$ have shown a reasonable fit at low and high temperatures with the lattice results. Therefore, this phenomenological model is more applicable to describe the EoS of the QGP matter rather than the Mayer's cluster expansion theory.

**References**

[1] T. Matsui and H. Satz, Phys. Lett. B 178,, 416 (1986).

[2] H. Satz, J. Phys. G 32, R25(2006), hep-ph/0512217.

[3] Nora Brambilla, Jacopo Ghiglieri, Peter Petreczky, Antonio Vairo, Phys. Rev. D78 , 014017 (2008), hep-ph/0804.0993.

[4] G. Boyd et al., Nucl. Phys. B 469, 419 (1996), hep-lat/9602007.

[5] Y. Maezawa et al., Phys. Rev. D 75, 074501 (2007), hep-lat/0702004.

[6] F. Karsch, E. Laermann, and A. Peikert, Nucl. Phys. B 605, 579 (2001), hep-lat/0012023.

[7] W.M. Alberico, A. Beraudo, A. De Pace and A. Molinari Phys. Rev. D72, 114011 (2005), hep-ph/0507084v1.

[8] E. V. Shuryak and I. Zahed, Phys. Rev. D 70, 054507 (2004), hep-ph/0403127.

[9] W. M. Alberico, A. Beraudo, A. De Pace, and A. Molinari, Phys. Rev. D 75, 074009 (2007), hep-ph/0609116.

[10] F. Brau, and F. Buisseret, Phys. Rev. C **76**, 065212 (2007).

[11] Boris A. Gelman1, 2, Edward V. Shuryak1 and Ismail Zahed, Phys. Rev. C74, 044909 (2006), nucl-th/0605046v1.

[12] J. Cleymans, R. V. Gavai and E. Suhonen, Phys. Rep. 130, 217 (1986).

[13] V. M. Bannur, Phys. Lett. B 362, 7 (1995).

[14] K. M. Udayanandan, P. Sethumadhavan, and V. M. Bannur, Phys. Rev. C 76, 044908 (2007).

[15]  M. Bluhm  B. Kampfer, R. Schulze, D. Seipt , Contribution to the International School of Nuclear Physics - 30th Course: Heavy-Ion Collisions from the Coulomb Barrier to the Quark-Gluon Plasma, Erice, Sicily, Italy, 16-24 Sep (2008),hep-ph/0901.0472v1.

[16]  L.M. Satarov, M.N. Dmitriev, I.N. Mishustin, Phys. Atom. Nucl.72, 1390 (2009), hep-ph/0901.1430v1.

[17]  Raghunath Sahoo, Tapan K. Nayak, Jan-e Alam, Sonia Kabana, Basanta K. Nandi, D. P. Mahapatra, Presented in SQM08, J. Phys. G36, 064071 (2009), nucl-ex/0901.3254v1.

[18]  E.V.Komarov, Yu.A.Simonov, talk given at "13$^{th}$ Lomonosov Conference on Elementary Particle Physics", Moscow, August 23 - 29, (2007), hep-ph/0801.2251v2.

[19] Ariel R. Zhitnitsky, Nuclear Physics A 813, 279 (2008).

[20] N.M. El Naggar, L. I. Abou-Salem, I. A. Elmashad, A.F. Ali, J. Mod. Phys.4, No. 4A,, 13 (2013).

[21] Michael Cheng, Proceedings of the XXV International Symposium on Lattice Field Theory, Regensburg, Germany (2007), hep-lat/0710.4357v1.

[22] B. Liu, P. N. Shen and H. C. Chiang, Phys. Rev. C55, No.6, 3021 (1997).

[23] Edward Shuryak, Prog.Part.Nucl.Phys.62, 48(2009), hep- ph/ 0807.3033v2.

[24] V.M. Bannur, J. Phys. G32, 993 (2006), hep-ph/0504072v2.

[25]  S. Ichimaru, Statistical Plasma Physics (Vol. II) - Condensed Plasma (Addison-Wesley Publishing Company, New York), (1994).

[26] V. M. Bannur, hep-ph /0807.2092v1.

[27] R. Rapp, D. Cabrera, V. Greco, M. Mannarelli and H. van Hees,  Winter Workshop on Nuclear Dynamics, South Padre Island (TX, USA), April 05-12 (2008), hep-ph/0806.3341v1.

[28] J. L. Nagle, Eur. Phys. J. C49, 275 (2007) , nucl-th/0608070 v1.

[29] N. O. Agasian and Yu. A. Simonov, Phys. Lett. B 639, 82 (2006), hep-ph/0604004.

[30] L. I. Abou-Salem, Int. J. Mod. Phys. A20, 4113 (2005),  L. I. Abou-Salem,  J. Phys. G: Nuclear and Particle Physics 30, 1391(2004).

[31] A. G. Tsypkin and G. G. Tsypkin, mathematical formulas, Mir publisher, Moscow (1988).

[32] S. M. Wong, computational methods in physics and engineering,© by Prentice – Hall, Inc. (1992).

[33] A. R. Gourlay, and G. A. Watson, computational methods for matrix eigenproblems, © John Wiely&Sons Ltd (1973).

[34] Jinfeng Liao and Edward Shuryak, Phys. Rev. C75, 054907 (2007).

[35] O. Portilho and Z.M.O. Shokranian, Revista Erasilieira de Flsica,Vol.14, No. 1, (1984).

[36] R. Balescu, Statistical Mechanics of Charged Particles © InterSceince Publishers, London (1963).

[37] D . Kremp, M. Schlanges and W. D. Kraeft, Quantum Statistics of Nonideal Plasmas@Springer-Verlag Berlin Heildberg (2005).

[38] J. Lestessier and J. Rafelski, hadrons and quark-gluon plasma © Syndicate of the University of Cambridge (2002).

[39] R. Abe, Progr. Theor. Phys. 21, 475 (1959).

[40] F. Karsch and E. Laermann, hep-lat/0305025 v1, "quark-gluon plasma III" Rudolph C. Hwa (Editor), X. N. Wang (Editor) (2004).

[41] G. Boyd, J. Engels, F. Karsch, E. Laermann, C. Legeland,M. Lütgemeier and B. Petersson, Phys.Rev. Lett. 75, 4169 (1995) and Nucl. Phys. B 469 419 (1996).

[42] J. Engles, F. Karsh, H. Satz, and I. Montavay, Nucl. Phys. B205, 545 (1982).